\pdfoutput=1
\documentclass[aps,preprint,superscriptaddress]{revtex4}
\usepackage{graphicx,amssymb,bm,natbib,amsfonts,amsmath}
\usepackage{hyperref}

\begin{document}

\title{Kohn anomaly and interplay of electron-electron and electron-phonon interactions in epitaxial graphene}

\author{S.Y. Zhou}
\affiliation{Department of Physics, University of California,
Berkeley, CA 94720, USA}
\affiliation{Materials Sciences Division,
Lawrence Berkeley National Laboratory, Berkeley, CA 94720, USA}

\author{D.A. Siegel}
\affiliation{Department of Physics, University of California,
Berkeley, CA 94720, USA}
\affiliation{Materials Sciences Division,
Lawrence Berkeley National Laboratory, Berkeley, CA 94720, USA}

\author{A.V. Fedorov}
\affiliation{Advanced Light Source, Lawrence Berkeley National Laboratory, Berkeley, California 94720, USA}

\author{A. Lanzara}
\affiliation{Department of Physics, University of California,
Berkeley, CA 94720, USA}
\affiliation{Materials Sciences Division,
Lawrence Berkeley National Laboratory, Berkeley, CA 94720, USA}

\date{\today}

\begin{abstract}
The interplay of electron-phonon (el-ph) and electron-electron (el-el) interactions in epitaxial graphene is studied by directly probing its electronic structure.  We found a strong coupling of electrons to the soft part of the A$_{1g}$ phonon evident by a kink at 150$\pm$15 meV, while the coupling of electrons to another expected phonon E$_{2g}$ at 195 meV can only be barely detected. The possible role of the el-el interaction to account for the enhanced coupling of electrons to the A$_{1g}$ phonon, and the contribution of el-ph interaction to the linear imaginary part of the self energy at high binding energy are also discussed.  Our results reveal the dominant role of the A$_{1g}$ phonon in the el-ph interaction in graphene, and highlight the important interplay of el-el and el-ph interactions in the self energy of graphene.
\end{abstract}

\maketitle

\begin{figure*}
\includegraphics[width=11.2 cm] {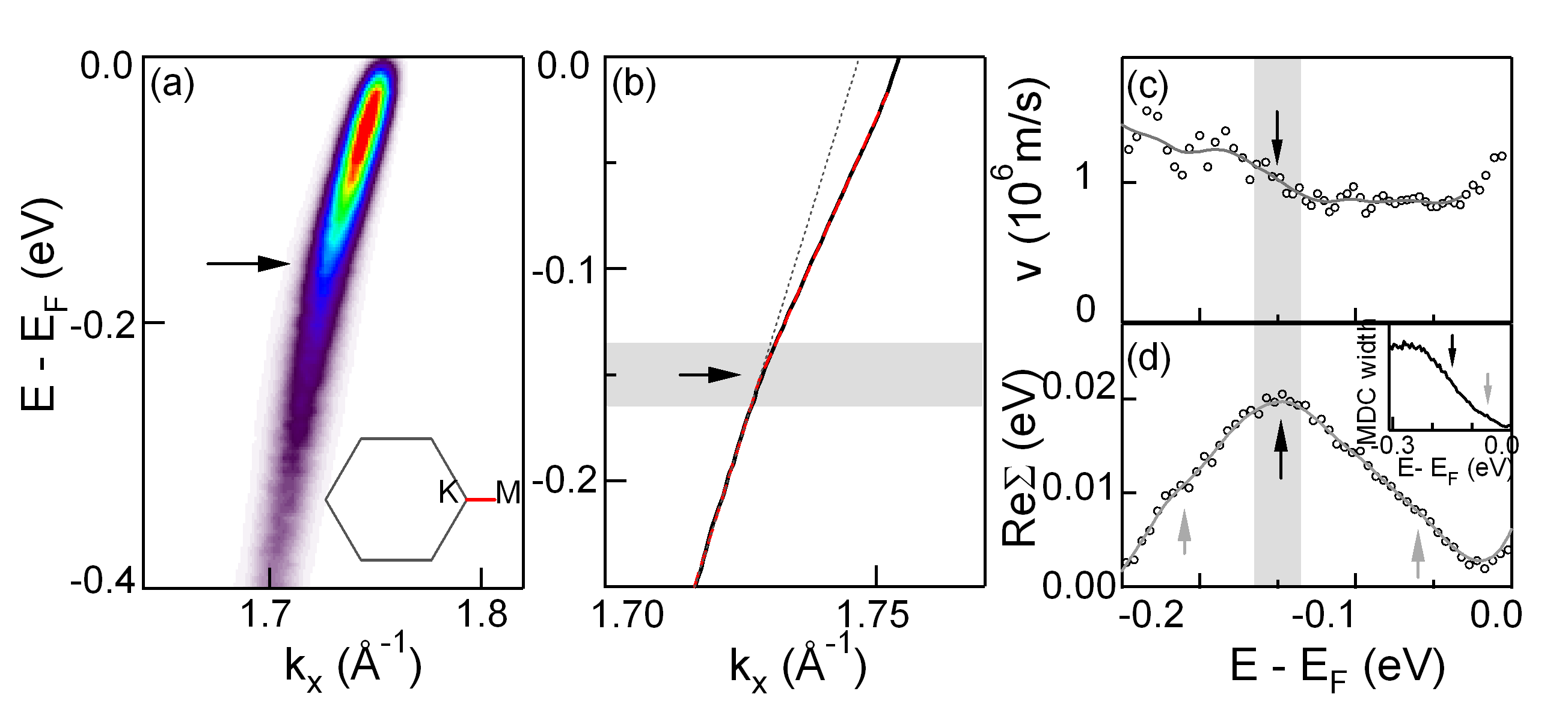}
\label{Figure 1}
\caption{(Color online) (a) ARPES data taken along KM direction (solid line in the inset).  (b) Dispersion (black curve) extracted by fitting the raw data.  The dashed line is the fit using two straight lines with different slopes. Within 20 meV below E$_F$, the dispersion is affected by the resolution, and therefore we fit the dispersion only in the range between -250 meV and -20 meV.  The gray dotted line is a guide for the deviation of the low energy dispersion from the extrapolation of the high energy dispersion.  (c) Extracted velocity as a function of energy from dispersions plotted in Panel b.  The symbols are the raw data and the black solid line is a guide for the eye.  (d) Extracted real part of the self energy as a function of energy Re$\Sigma$(E)=E(k)-v$_{b}$k$_F$ where E(k) is the measured dispersion and k$_F$ is the Fermi wave vector.  The inset shows the MDC width.}
\end{figure*}

Electron-phonon (el-ph) coupling is among the most important interactions, since it is at the origin of a variety of interesting phenomena, such as the hopping-like charge transport in organic semiconductors \cite{ElphGas, ElphSolid}, charge density wave formation \cite{Gruner}, metal-insulator transition, superconductivity \cite{BCS} and ballistic transport \cite{Javey1, Yao}.  The el-ph interaction is particularly intriguing in graphitic materials, where the special electronic properties of Dirac fermions and the interplay of el-ph and electron-electron (el-el) interaction result in a wide range of novel physics \cite{NetoRMP}.  Because of its peculiar point-like Fermi surface, which can be connected by the wave vectors of the phonons at $\Gamma$ and K \cite{PhononDisp}, electron screening of the lattice vibrations decreases dramatically around these two points, causing two Kohn anomalies \cite{KohnAnomaly}.  Moreover in the case of single layer graphene, the peculiar band structure also results in the breakdown of the Born Oppenheimer approximation \cite{Mauri, AHCN};  
a shift of the E$_{2g}$ phonon frequency as a function of carrier concentration \cite{AHCNPRB, AHCN, Mauri, Yan} and sample thickness \cite{Eklund}; and a predicted anomalous phonon-induced self energy \cite{Calandra, Louieelph, Sarma} that deviates from that of conventional metals \cite{ME}. 

Despite the intense research effort, two key components in understanding el-ph interaction - the phonon modes involved and the coupling strength of the interaction, have not been resolved.  Angle-resolved photoemission spectroscopy (ARPES) is an ideal tool in this respect, as it directly measures the renormalized electronic band structure of a material and therefore provides direct insights about many-body interactions.  In recent years ARPES has been successfully used to detect the signature of the el-ph interaction in the electronic spectra in the form of a kink, in both graphite \cite{AnnalsPhys, Sugawara, Kim} and graphene \cite{EliNatPhys, Jessica}.  However, not only is there a discrepancy in the value of the observed coupling strength \cite {AnnalsPhys, Sugawara, EliNatPhys, Kim, Jessica} with respect to the theoretical predictions \cite{Calandra, Louieelph, Sarma},  but also consensus on which and how many phonon modes are involved has been missing so far.  Theoretically it was proposed that due to the Kohn anomalies at $\Gamma$ and K, both the E$_{2g}$ (195 meV) and A$_{1g}$ (165 meV) phonons contribute to the el-ph interaction \cite{Calandra}.  Experimentally although a kink has been reported in the electronic dispersion \cite{EliNatPhys, Jessica}, the large uncertainty in the kink energy makes it difficult to distinguish which, if not both, phonons are involved.  Therefore a more detailed study with improved data quality is needed to complete our understanding of the el-ph coupling in graphene and provide key insights for the el-ph coupling in other graphitic materials.

In this letter we present a high resolution ARPES study of the el-ph interaction and its contribution to the electron self energy in epitaxial graphene.  The greatly improved data quality with reduced noise level has enabled us for the first time to nail down the kink energy in the electronic dispersion to 150$\pm$15 meV and to reveal additional fine structures in the electron self energy at 60$\pm$15 meV and 200$\pm$15 meV.  More importantly, the direct comparison between the electronic dispersion measured here and the reported phonon dispersion relation \cite{PhononDisp} has allowed us to identify the soft part of the A$_{1g}$ phonon (Kohn anomaly) as the main scattering channel responsible for the ARPES kink, and the fine structure at 200 meV in the self energy with the E$_{2g}$ mode.  The enhanced coupling to the A$_{1g}$ mode with respect to the E$_{2g}$ mode together with the much larger experimental el-ph coupling strength $\lambda$$\approx$0.14 as compared to theoretical one is discussed in terms of Coulomb interactions.  In addition, we report the linear imaginary part of the self energy at high binding energy with similar magnitude along various directions, which reflects the contribution from both el-ph and el-el interactions.  Our results point to the dominant role of A$_{1g}$ phonon in the el-ph interaction in graphene, and highlight the important interplay of el-ph and el-el interaction in the intriguing physics of Dirac fermions in graphene. 

High resolution ARPES data were taken on single layer epitaxial graphene at Beamline 12.0.1 (Figs.1-3) and Beamline 7.0.1 (Fig.~4) of the Advanced Light Source (ALS) of the Lawrence Berkeley National Laboratory with a total energy resolution of 25 meV and 35 meV respectively.  Samples were grown on a n-type SiC wafers as detailed elsewhere \cite{BergerJPC, Liz}.  The samples were measured with 50 eV photon energy at a temperature of 25K, with vacuum better than 3.0$\times$10$^{-11}$ Torr. 

\begin{figure}
\includegraphics [width=7.2 cm] {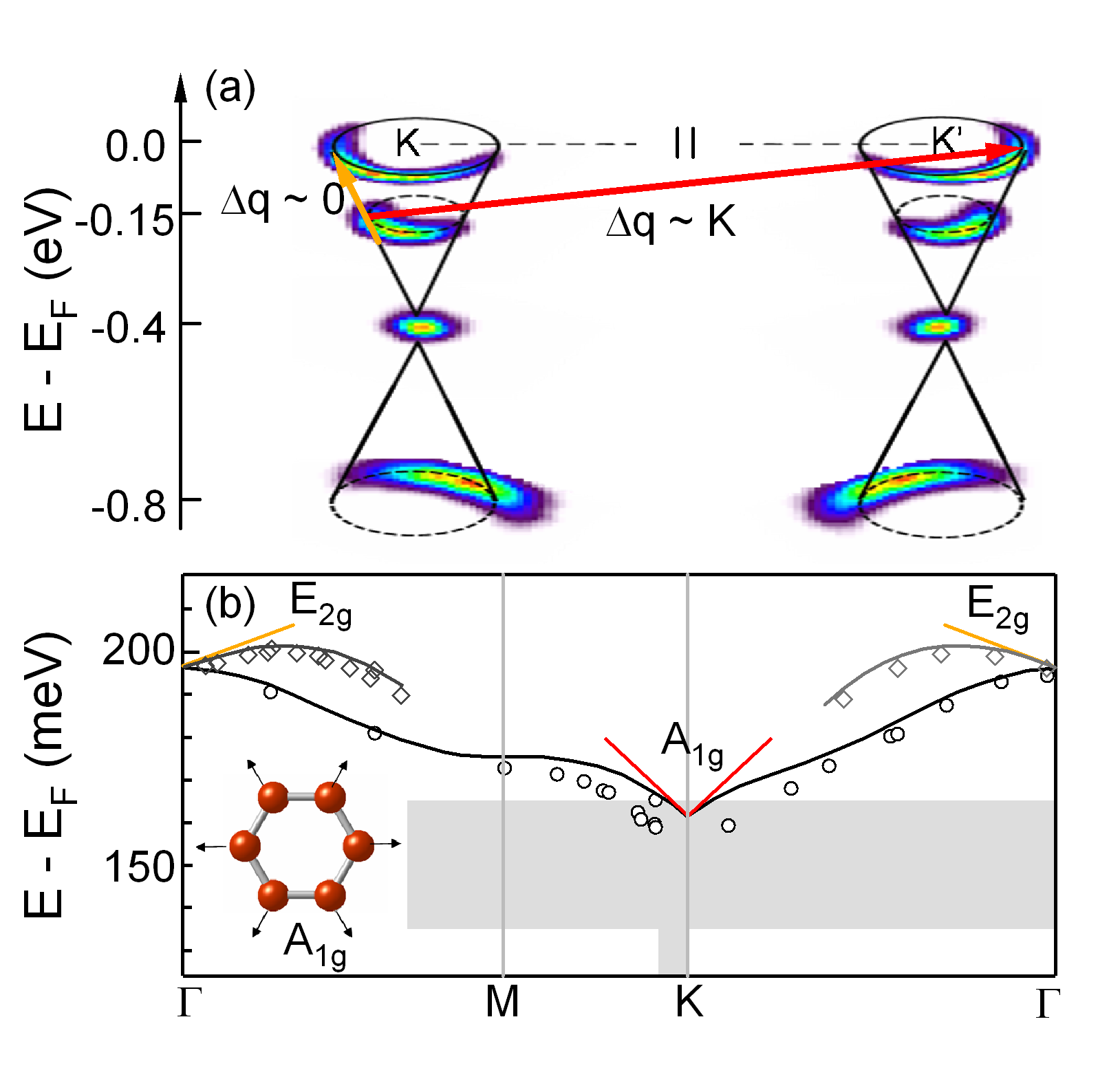}
\label{Figure 2}
\caption{(color online) (a) Intensity maps for the neighboring K and K$^\prime$ points as a function of energy and momentum. The two arrows show the el-ph interaction with the E$_{2g}$ phonon and A$_{1g}$ phonon respectively.  (b) Phonon dispersions for the A$_{1g}$ near K and E$_{2g}$ near $\Gamma$ \cite{PhononDisp,KohnAnomaly}.}
\end{figure}

\begin{figure*}
\includegraphics [width=11.2 cm] {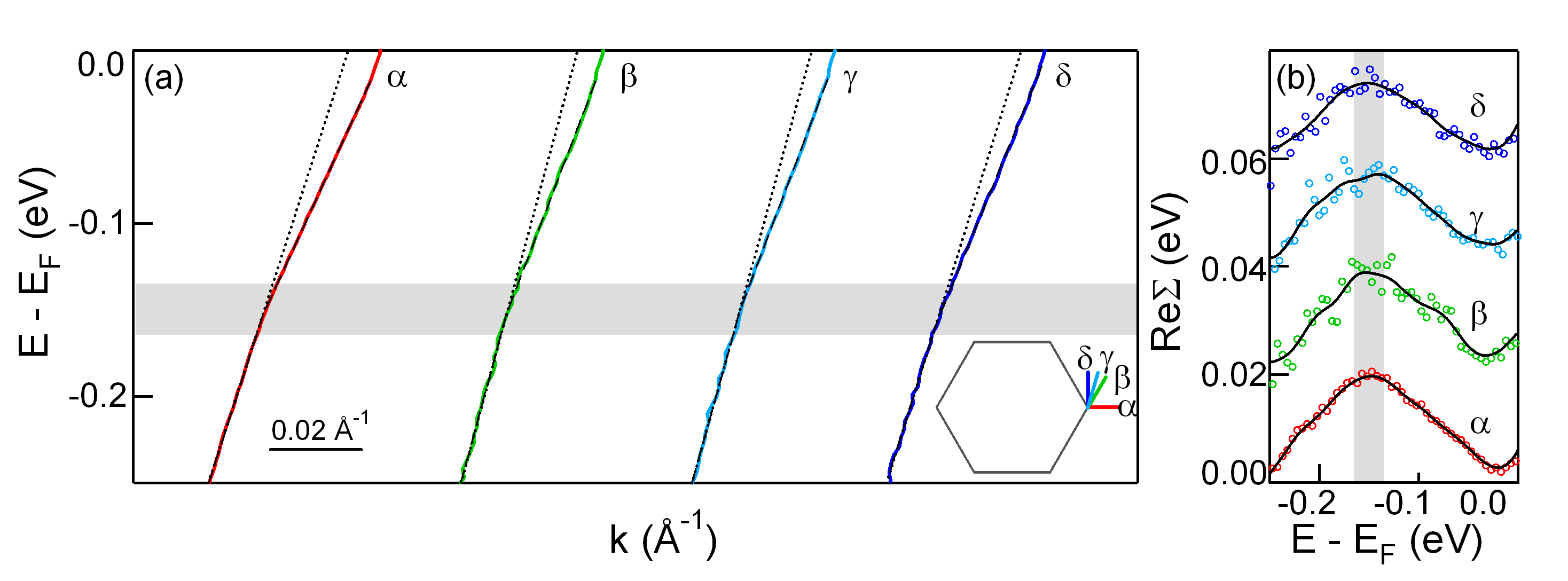}
\label{Figure 3}
\caption{(Color online) (a) Dispersions along various directions (labeled as $\alpha$, $\beta$, $\gamma$ and $\delta$) shown in the Brillouin zone in the inset.  The dashed lines are fits of the dispersion using two straight lines with different slopes.  The dotted line is a guide to the eye for the deviation at low binding energy.  The horizonal gray shadow highlights the kink energy. (b) Corresponding Re$\Sigma$ extracted by subtracting the bare band dispersion from the measured dispersion.  The bare band dispersion is taken as a straight line connecting the dispersions at E$_F$ and -250 meV.}
\end{figure*}
 
\begin{figure}
\includegraphics [width=7.4 cm] {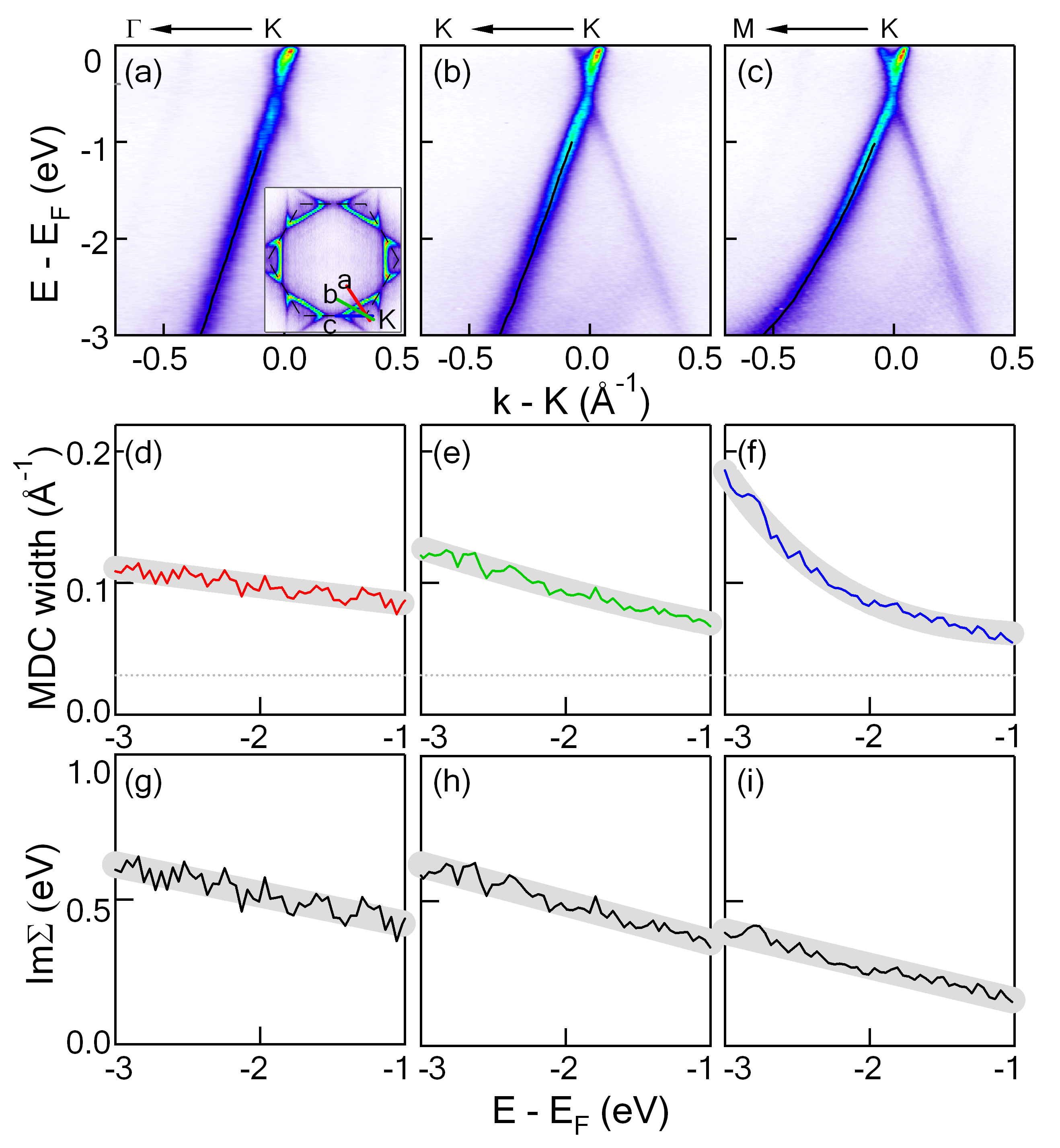}
\label{Figure 4}
\caption{(Color online) (a-c) Dispersions along various directions. The inset shows the constant energy map at -3 eV, where the trigonal distortion can be clearly observed.  The three lines in the inset label the three cuts shown in panels a-c. (d-f) Extracted MDC width as a function of energy for data shown in panels a-c.  The gray line is a guide for the transition from a more linear behavior to a more quadratic behavior.  (g-i) Imaginary part of the self energy Im$\Sigma$=$\frac{1}{2}$v$\cdot$$\Delta$k, where v is the velocity at each energy and $\Delta$k is the MDC width subtracted by the width at E$_F$ (dotted lines in panels d-f) to take care of the impurity scattering and finite experimental resolution.}
\end{figure}

Figure 1a shows an ARPES intensity map taken through the Dirac point (K point). One can easily identify a characteristic energy (pointed to by a horizontal black arrow), where the slope of the dispersion (i.e.~velocity) changes and the intensity suddenly decreases due to the disappearance of coherent peaks in the energy distribution curves (EDCs) at high binding energy.  These are typical signatures of electron-boson coupling, where the broadening of the spectra beyond the kink energy is due to the onset of the bosonic mode self energy \cite{AnnalsPhys, Sugawara, Kim, EliNatPhys, Jessica}.  
To identify the exact bosic modes involved in the coupling and the strength of the coupling, we extract the dispersion relation from the peak positions and the Im$\Sigma$ from the peak width by fitting the momentum distribution curves (MDCs).  
The high statistics of the data in panel b allows us to nail down the kink position to 150$\pm$15 meV \cite{note}.  The extracted kink energy is also consistent with a sudden drop of the MDC width (inset of panel d) and a change of the electron velocity (panel c), both occurring at -150 meV.  From the renormalization of the electron velocity we extract the el-ph coupling constant $\lambda$ = v$_b$/v$_F$-1, where v$_b$=1.0$\times$10$^6$m/s is the bare band velocity, and v$_F$ the renormalized Fermi velocity v$_F$=0.87$\times$10$^6$m/s.  This gives an experimental $\lambda$ = 0.14 which is almost an order of magnitude larger than the predicted value of 0.02 for the A$_{1g}$ phonon \cite{Calandra}.  The identification of the kink at 150 meV with strength of 0.14 is also supported by recent scanning tunneling microscope measurements \cite{Andrei}.

To check whether other phonon modes contribute to this large value of $\lambda$, we show in panel d the real part of the electron self energy, Re$\Sigma$.  In addition to the main peak at -150 meV that dominates Re$\Sigma$, two additional fine structures at $\approx$ -60 meV and $\approx$ -200 meV can also be resolved. 
The existence of these fine structures indicates the involvement of other collective modes in the coupling. 
Since the area underneath the Re$\Sigma$ is an indication of the coupling strength, clearly these additional modes contribute only a small fraction to the total coupling constant, and hence cannot be responsible for the large discrepancy between the experimental and theoretical $\lambda$.

To single out the allowed scattering processes, in figure 2 we compare the ARPES constant energy map at E$_F$ and at the kink energy (panel a) with the predicted phonon dispersion (panel b) \cite {PhononDisp}. 
Clearly the soft part of the A$_{1g}$ phonon near the zone corner K point is the only mode with the right energy and momentum q$\approx$$|\Gamma K|$ to scatter states separated in energy by 150 meV (kink energy) from near the K point to the K$^\prime$ point (inter-valley scattering), therefore being likely the dominant source for the kink in the dispersion and the maximum peak in the Re$\Sigma$.  Similarly, the E$_{2g}$ phonon near the $\Gamma$ point has the right energy (195 meV) and momentum q$\approx$0 to connect states between 200 meV and the Fermi energy within the same K point (intra-valley scattering) and is responsible for the fine structure in Re$\Sigma$ at 200 meV.  Since the sample is slightly electron doped, the q vector for intra- and inter-valley scattering is $\approx$4$\%$$|\Gamma K|$ larger than q=$|\Gamma K|$ and q=0, but still in the proximity of the two Kohn anomalies \cite{KohnAnomaly}.  Finally, the fine structure in Re$\Sigma$ at 60 meV is likely due to coupling with an out-of-plane phonon as reported by STM studies \cite{Crommie}. 

Figure 3 compares the el-ph interaction along different directions.  
A similar kink is present in the dispersion along all the directions at the same energy of 150 $\pm$ 15 meV (see gray region in panel a).  
Although similar additional fine structures involving the two other phonon modes are also observed in the self energy in panel b, the most important finding is that the  A$_{1g}$ phonon is still the dominant one.

Although on a qualitative level the data presented here are in good agreement with theoretical calculations, on a quantitative level there are two key differences. 1) Experimentally we found that the intervalley scattering with the A$_{1g}$ phonon is by far the {\it dominant} scattering source.  This is in contrast to theoretical prediction where both intervalley (A$_{1g}$ phonon) and intravalley (E$_{2g}$ phonon) scatterings are treated in an almost equal footing, although the latter is decreased by half \cite{KohnAnomaly}.  2) The experimental el-ph coupling strength of 0.14, mostly accounted for by the A$_{1g}$ phonon (as discussed in figure 1),  is much larger than the predicted value of 0.02 \cite{Calandra}.  This holds even if finite experimental resolution, which makes $\lambda$ twice as big, is taken into account \cite{Calandra}.  Therefore, additional mechanism needs to be included to explain the observed enhancement, by approximately a factor of 3, of the el-ph coupling strength.  
One likely candidate is through the interplay with el-el interaction, as  pointed out theoretically \cite {Calandra, Louieelph}.  More specifically, it has been argued that this interplay gives rise, in the presence of a linear dispersion, to a linear imaginary part of the self energy Im$\Sigma$ in agreement with experimental reports \cite{EliNatPhys, Xu}, and that the el-ph coupling contributes to 1/3 of its total magnitude \cite {Louieelph}.  By extending this study to the entire momentum region (Figure 4), we shown that this linearity in Im$\Sigma$ survives with a similar magnitude throughout the entire Dirac cone, even when the dispersion is not linear because of the trigonal distortions (Fig.4(b,c)),  and is hence a general property of Dirac fermions. 
Panels (a-c) show the ARPES intensity maps along three different directions.  From $\Gamma$K (panel a) to MK direction (panel c), both the extracted dispersion (see solid black line in panels a-c) and MDC width between -1 and -3 eV change from a linear to a quadratic behavior, due to the trigonal distortion (see also inset of panel a) \cite{ZhouNatPhys}.  The corresponding Im$\Sigma$ are shown in panels (g-i).  Clearly, even when the dispersion is not linear (panel c), Im$\Sigma$ still shows a linear dependence with {\itshape similar} magnitude (panel i), suggesting that the overall contribution of the el-ph interaction to the self energy of graphene is comparable along all directions, in line with the isotropic el-ph coupling reported in Fig.~3 and theoretical prediction \cite{Parkelph}. 

These data clearly establish the importance of the interpaly between el-ph and el-el interaction suggesting that the latter might be responsible for the observed enhancement of the coupling strength.  Indeed it has been argued that in the completely unscreened case, the el-el interaction can enhnace the coupling to the A$_{1g}$ phonon near K by up to a factor of 3, leaving the coupling to the E$_{2g}$ phonon near $\Gamma$ almost unaffected \cite{Basko}.  This picture reconciles the disagreement between the theoretical and experimental coupling strength measured by ARPES and can also account for the large intensity ratio between the 2D and 2D$^\prime$ peaks reported by Raman \cite{Ferrari}, which is likely due to the enhanced renormalization of the el-ph coupling of peak D.  Finally it is interesting to note that a similar enhanced coupling to phonons with non-zero wave vectors through el-el interaction also occurs in the case of transition metal dichalcogenides where the quasiparticles are Dirac fermions, resulting as well in a linear Im$\Sigma$ \cite{NetoPRL01}. These similarities suggest that the physics here discussed is not only a property of graphene but a more general property of Dirac materials.

In conclusion, we have reported the strong interplay of el-ph and el-el interactions in graphene.  We identified the dominant role of the A$_{1g}$ phonons at 150 meV in the el-ph interaction along the various directions near the K point.  Although the fine structures due to coupling with other phonon modes are observed, we show that the enhancement of the coupling to the soft part of the A$_{1g}$ mode is likely induced by the interplay between el-el and el-ph interactions.  This study demonstrates the important role of this interplay in a Dirac fermion system, and highlights the importance of including both interactions in the self energy of graphene.

\begin{acknowledgments}
We thank D.-H. Lee for useful discussions. This work was supported by the National Science Foundation through Grant No.~DMR03-49361 and the Director, Office of Science, Office of Basic Energy Sciences, Division of Materials Sciences and Engineering of the U.S Department of Energy under Contract No.~DEAC03-76SF00098 and and by the Laboratory Directed Research and Development Program of Lawrence Berkeley National Laboratory under the Department of Energy Contract No. DE-AC02-05CH11231.
S.Y. Zhou thanks the Advanced Light Source Fellowship for financial support.
\end{acknowledgments}



\begin {thebibliography} {99}

\bibitem{ElphGas} V. Coropceanu, M. Malagoli, D.A. da Silva Filho, N.E. Gruhn, T.G. Bill and J.L. Bredas, Phys. Rev. Lett. {\bf 89}, 275503 (2002).

\bibitem{ElphSolid} H. Yamane, S. Nagamatsu, H. FUkagawa, S. Kera, R. Friedlein, K.K. Okudaira and N. Ueno, Phys. Rev. B {\bf 72}, 153412 (2005).

\bibitem{Gruner} G. Gruner,  Rev. Mod. Phys. {\bf 60}, 1129 (1988).

\bibitem{BCS} J. Bardeen, L.N. Cooper and J.R. Schrieffer, Phys. Rev. {\bf 108}, 1175 (1957).

\bibitem{Javey1} A. Javey, J. Guo, Q. Qang, M. Lundstrom and H. Dai, Nature, {\bf 424}, 654 (2003).

\bibitem{Yao} Z. Yao, C.L. Kane and C. Dekker, Phys. Rev. Lett. {\bf 84}, 2941 (2000).

\bibitem{NetoRMP} A.H. Castro Neto, F. Guinea, N.M.R. Peres, K.S. Novoselov and A.K. Geim,  arXiv:0709.1163 (2008). %

\bibitem{PhononDisp} J. Maultzsch, S. Reich, C. Thomsen, H. Requardt and P. Ordejon, Phys. Rev. Lett. {\bf 92}, 075501 (2004).

\bibitem{KohnAnomaly} S. Piscanec, M. Lazzeri, F. Mauri, A.C. Ferrari and J. Robertson,  Phys. Rev. Lett. {\bf 93}, 185503 (2004).

\bibitem{Mauri} S. Pisana, M. Lazzeri, C. Casiraghi, K.S. Novoselov, A.K. Geim, A.C. Ferrari and F. Mauri, Nature Mat. {\bf 6}, 198 (2007).

\bibitem{AHCN} A.~H. Castro Neto, Nature Mat. {\bf 6}, 176 (2007).

\bibitem{Yan} J. Yan, Y. Zhang, P. Kim and A. Pinczuk, Phys. Rev. Lett. {\bf 98}, 166802 (2007).

\bibitem{AHCNPRB} A.H. Castro Neto and F. Guinea, Phys. Rev. B {\bf 75}, 045404 (2007).

\bibitem{Eklund} A. Gupta, G. Chen, P. Joshi, S. Tadigadapa and P.C. Eklund, Nano Lett. {\bf 6}, 2667 (2006).

\bibitem{Calandra} M. Calandra and F. Mauri, Phys. Rev. B {\bf 76}, 205411 (2007).

\bibitem{Louieelph} C.-H. Park, F. Giustino, M.L. Cohen and S.G. Louie, Phys. Rev. Lett. {\bf 99}, 086804 (2007).

\bibitem{Sarma} W.-K. Tse and S. Das Sarma, Phys. Rev. Lett. {\bf 99}, 236802 (2007).

\bibitem{ME} Ashcroft and Mermin, {\it Solid State Physics}.

\bibitem{AnnalsPhys} S.Y. Zhou, G.-H. Gweon and A. Lanzara, Annals of Physics, {\bf 321}, 1730 (2006).

\bibitem{Sugawara} K. Sugawara, T. Sato, S. Souma, T. Takahashi and H. Suematsu, Phys. Rev. Lett., {\bf 98}, 036801 (2007).

\bibitem{Kim} C.S. Leem, B.J. Kim, S.R. Park, T. Ohta, A. Bostwick, E. Rotenberg, H.-D. Kim, M.K. Kim, H.J. Choi and C. Kim, Phys. Rev. Lett. {\bf 100}, 016802 (2008).

\bibitem{EliNatPhys} A. Bostwick, T. Ohta, T. Seyller, K. Horn and E. Rotenberg, Nature Phys., {\bf 3}, 36 (2006).

\bibitem{Jessica} J.L. McChesney, A. Bostwick, T. Ohta, K.V. Emtsev, Th. Seyller, K. Horn and E. Rotenberg, arXiv:0705.3264 (2007).

\bibitem{BergerJPC} C. Berger, Z. Song, T. Li, X. Li, A.Y. Ogbazghi, R. Feng, Z. Dai, A.N. Marchenkov, E.H. Conrad, P.N. First and W.A. de Heer, J. Phys. Chem. B {\bf 108}, 19912 (2004).

\bibitem{Liz} E. Rollings, G.-H. Gweon, S.Y. Zhou, B.S. Mun, J.L. McChesney, B.S. Hussain, A.V. Fedorov, P.N. First, W.A. de Heer and A. Lanzara,  J. Phys. Chem. Solids {\bf 67}, 2172 (2006). 

\bibitem{note} Note that the kink energy is far away from the deviation near the Dirac point ($\approx$ -0.4 eV) reported in epitaxial graphene \cite{EliNatPhys,ZhouNatMat}, and therefore is unaffected by such deviation. 

\bibitem{Andrei} G. Li, A. Luican, E.Y. Andrei.  arXiv:cond-mat/0803.4016 (2008).

\bibitem{Crommie} Y.B. Zhang, V.W. Brar, F. Wang, C. Girit, Y. Yayon, M. Panlasigui, A. Zettl and M.F. Crommie, Nature Phys. {\bf 4}, 627 (2008).

\bibitem{Xu} S. Xu, J. Cao, C.C. Miller, D.A. Mantell, R.J.D. Miller and Y. Gao, Phys. Rev. Lett. {\bf 76}, 483 (1996).

\bibitem{ZhouNatPhys} S.Y. Zhou, G.-H. Gweon, J. Graf, A.V. Fedorov, C.D. Spataru, R.D. Diehl, Y. Kopelevich, D.-H. Lee, S.G. Louie and A. Lanzara, Nature Phys. {\bf 2}, 595 (2006).

\bibitem{Parkelph} C.-H. Park, F. Guistino, J.L. McChesney, A. Bostwick, T. Ohta, E. Rotenberg, M.L. Cohen and S.G. Louie, Phys. Rev. B {\bf 77}, 113410 (2008).

\bibitem{Basko} D.~M. Basko and I.L. Aleiner, Phys. Rev. B {\bf 77}, 041409(R) (2008).


\bibitem{Ferrari} A.C. Ferrari, J.C. Meyer, V. Scardaci, C. Casiraghi, M. Lazzeri, F. Mauri, S. Piscanec, D. Jiang, K.S. Novoselov, S. Roth and A.K. Geim, Phys. Rev. Lett. {\bf 97}, 187401 (2006).

\bibitem{NetoPRL01} A.H. Castro Neto, Phys. Rev. Lett. {\bf 86}, 4382 (2001).

\bibitem{ZhouNatMat} S.Y. Zhou, G.-H. Gweon, A.V. Fedorov, P.N. First, W.A. de Heer, D.-H. Lee, F. Guinea, A.H. Castro Neto and A. Lanzara, Nature Mat. {\bf 6}, 770 (2007).



\end {thebibliography}
\end{document}